\documentclass[12pt,fullpage,doublespace,epsf]{article}
\usepackage{amsmath}
\usepackage{amssymb}
\usepackage{graphicx}
\numberwithin{equation}{section}
\DeclareMathSizes{11}{19}{13}{9}
\makeatletter
\@addtoreset{equation}{section}
\makeatother

\newcommand\BBN{\mbox{\tiny{BBN}}}

\newcommand\G{\mbox{G}}

\newcommand\eV{\mbox{eV}}

\newcommand\MeV{\mbox{MeV}}

\newcommand\dec{\mbox{\scriptsize dec}}
\newcommand\eq{\mbox{\scriptsize eq}}

\begin{document}

\begin{center}
{\bf Cosmic Microwave Background Quadrupole and Ellipsoidal Universe}
\end{center}

\begin{center}
L. Campanelli$^{a,b,}$\footnote{E-mail: campanelli@fe.infn.it},
P. Cea$^{c,d,}$\footnote{E-mail: paolo.cea@ba.infn.it}, and
L. Tedesco$^{c,d,}$\footnote{E-mail: luigi.tedesco@ba.infn.it}
\end{center}

\begin{center}
$^{a}${\em INFN - Sezione di Ferrara, I-44100 Ferrara, Italy}, \\
$^{b}${\em Dipartimento di Fisica, Universit\`a di Ferrara, I-44100 Ferrara, Italy}, \\
$^{c}${\em INFN - Sezione di Bari, I-70126 Bari, Italy}, \\
$^{d}${\em Dipartimento di Fisica, Universit\`a di Bari, I-70126 Bari, Italy}
\end{center}

\vspace{1.0cm}


\begin{abstract}
Recent Wilkinson Microwave Anisotropy Probe (WMAP) data confirm
the Cosmic Microwave Background (CMB) quadrupole anomaly. We
further elaborate our previous proposal that the quadrupole power
can be naturally suppressed in axis-symmetric universes. In
particular, we discuss in greater detail the CMB quadrupole
anisotropy and considerably improve our analysis. As a result, we
obtain tighter constraints on the direction of the axis of
symmetry as well as on the eccentricity at decoupling. We find
that the quadrupole amplitude can be brought in accordance with
observations with an eccentricity at decoupling of about $0.64
\times 10^{-2}$. Moreover, our determination of the direction of
the symmetry axis is in reasonable agreement with recent
statistical analyses of cleaned CMB temperature fluctuation maps
obtained by means of improved internal linear combination methods
as Galactic foreground subtraction technique.
\end{abstract}


\newpage

\section{\normalsize{Introduction}}
\renewcommand{\thesection}{\arabic{section}}

\noindent The Cosmic Microwave Background (CMB) has had a profound
impact on modern cosmology and greatly improved our understanding
of the universe. The CMB angular power spectrum is indeed very
sensitive to the origin and evolution of the cosmic density fluctuations.  \\
\indent
The temperature fluctuations of CMB are observed at the level of
$\Delta T / \langle T \rangle \sim 10^{-5}$~\cite{Smoot}. The high
resolution data provided by the Wilkinson Microwave Anisotropy
Probe (WMAP)~\cite{WMAPsite,Spergel,1yearWMAP,WILC3YR} confirmed
that the CMB anisotropy data are in striking agreement
with the predictions of the simplest inflation model. \\
\indent
However, the 3-years WMAP data (WMAP3) display at large angular
scales some anomalous features. The most important discrepancy
resides in the low quadrupole moment, which signals an important
suppression of power at large scales, although the probability of
quadrupole being low is not statistically compelling. The problem
consists into the fact that the power of the quadrupole is
substantially reduced with respect to the value of the best-fit
$\Lambda$-dominated cold dark matter ($\Lambda$CDM) standard
model. If this discrepancy turns out to have a cosmological
origin, then it could have far reaching consequences for our
understanding of the universe and in particular for the standard
inflationary picture. This peculiarity emerged since 1992 when the
first data of the differential microwave radiometer (COBE/DMR)
appeared~\cite{Smoot}. Since then, in 2003 (WMAP) and in 2006
(WMAP3) this behavior was confirmed. In fact, the WMAP3 data give
a quadrupole power of 211 $\mu\mbox{K}^2$, while the expected
value in $\Lambda$-dominated cold dark matter model is
about 1252 $\mu\mbox{K}^2$. \\
\indent
In the last years, the ``smallness'' of CMB quadrupole has been
subject to very intensive studies because it may signal a
non-trivial topology of the large scale geometry of the
universe~\cite{Topology}. Indeed, several possibilities have been
advanced in the recent literature to understand the suppression of
the quadrupole power~\cite{Quadrupole1,Quadrupole2} (for other
large scale anomalies in the angular distribution of CMB see
Ref.~\cite{Anomalies}).
\\
\indent
Recently~\cite{prl} we showed that, allowing the large-scale
spatial geometry of our universe to be plane-symmetric with
eccentricity at decoupling of order $10^{-2}$, the quadrupole
amplitude could be drastically reduced without affecting higher
multipoles of the angular power spectrum of the temperature
anisotropy. Remarkably, the ``ellipsoidal'' (or ``eccentric'')
universe has been considered also in non-standard cosmological
models, such as braneworld cosmology~\cite{Ge}.
\\
There are different mechanisms which could induce a planar
symmetry in the spatial geometry of the universe~\cite{Berera}.
Among these, the most interesting examples include a cosmic domain
wall, a cosmic string and an almost uniform cosmic magnetic field.
In particular, the cosmic magnetic field seems to be of relevance
since several observations suggest the presence of magnetic fields
correlated on cosmic scales~\cite{CMF} (for a recent study of the
effects of cosmic magnetic fields on large-scale structures of the
universe see Ref.~\cite{Barrow}, while for influence of cosmic
fields on the expansion of the universe see
Ref.~\cite{Matravers}). There are several possible mechanisms to
generate cosmological magnetic fields. For instance, magnetic
fields might be produced during~\cite{Generation1,Generation2} or
after~\cite{Generation3} the inflation era. \\
\indent
In this paper, we further discuss the proposal advanced in
Ref.~\cite{prl} where we showed that cosmic microwave background
quadrupole power can be naturally suppressed in plane-symmetric
universes. In particular, we discuss in greater detail the CMB
quadrupole anisotropy and considerably improve the analysis
presented in our previous paper~\cite{prl}. As a result, we obtain
tighter constraints on the direction of the axis
of symmetry as well as on the eccentricity at decoupling. \\
\indent
The plan of the paper is as follows. In Sec.~2 we discuss the
Einstein's equations for cosmological models with planar geometry,
and we describe the mechanisms to generate the eccentricity in the
universe expansion by means of magnetic fields, domain walls or
cosmic strings; Sec.~3 deals with the analysis of CMB anisotropies
including the asymmetric contributions to the temperature
anisotropy. In Sec.~4 we discuss the constraints on cosmic
magnetic fields arising from primordial nucleosynthesis the large
scale structure formation. Finally, we draw our conclusions in
Sec.~5. Some technical details are relegated in the Appendix.

\section{\normalsize{Cosmological models with planar symmetry}}
\renewcommand{\thesection}{\arabic{section}}

\noindent We are interested in cosmological models with planar
symmetry. The most general plane-symmetric line
element~\cite{Taub} is:
\begin{equation}
\label{eq1}
ds^2 = dt^2 - a^2(t) (dx^2 + dy^2) - b^2(t) \, dz^2 ,
\end{equation}
where $a$ and $b$ are the scale factors. The metric (\ref{eq1})
corresponds to considering the $xy$-plane as a symmetry plane.
\\
The non-zero Christoffel symbols corresponding to the metric
(\ref{eq1}) are:
\begin{equation}
\label{eq2}
\Gamma^{0}_{11} = \Gamma^{0}_{22} = a \dot{a}, ~~ \Gamma^{0}_{33}
= b \dot{b}, ~~ \Gamma^{1}_{01} = \Gamma^{2}_{02} = \dot{a}/a, ~~
\Gamma^{3}_{03} = \dot{b}/b,
\end{equation}
where a dot indicates the derivative with respect to the cosmic
time. The non-zero Ricci tensor components turn out to be:
\begin{eqnarray}
\label{eq3}
R^0_0 & = & -2 \left(\frac{\ddot{a}}{a} +
\frac{\ddot{b}}{b} \right),
\\
R^1_1 & = & R^2_2  =  - \left[\frac{\ddot{a}}{a}+ \left(\frac
{\dot{a}} {a} \right)^2 + \frac{\dot{a}}{a} \frac{\dot{b}}{b}
\right],
\\
R^3_3 & = & - \left( \frac{\ddot{b}}{b} + 2\, \frac{\dot{a}}{a}
\frac{\dot{b}}{b} \right).
\end{eqnarray}
The most general energy-momentum tensor consistent with planar
symmetry is
\begin{equation}
\label{tensor}
T^{\mu}_{\,\,\, \nu} = \mbox{diag} \,
(\rho,-p_{\|},-p_{\|},-p_{\bot}).
\end{equation}
It can be made up of two different components: an anisotropic
contribution,
\begin{equation}
\label{eq4}
{(T_A)}^{\mu}_{\;\; \nu} = \mbox{diag} \,
(\rho^A,-p^A_{\|},-p^A_{\|},-p^A_{\bot}),
\end{equation}
which induces the planar symmetry --as, for example, a uniform
magnetic field, a domain wall, or a cosmic string--, and an
isotropic contribution,
\begin{equation}
\label{eq4a}
{(T_I)}^{\mu}_{\;\; \nu} = \mbox{diag} \, (\rho^I,-p^I, -p^I,
-p^I),
\end{equation}
such as vacuum energy, radiation, matter, or cosmological
constant. Exact solutions of Einstein's equations for different
kind of plane-symmetric plus isotropic components can be found in
Ref.~\cite{Berera}.
\\
Taking into account the above energy-momentum tensors, the
Einstein's equations
\begin{equation}
\label{eq4b}
R_{\mu \nu} - \frac{1}{2} g_{\mu \nu} R = 8\pi G T_{\mu \nu},
\end{equation}
read
\begin{eqnarray}
\label{eq5}
&& \left( \frac{\dot{a}}{a} \right)^{\!2} + 2 \,
\frac{\dot{a}}{a} \frac{\dot{b}}{b} = 8 \pi G (\rho^I+\rho^A), \\
\label{eq6}
&& \frac{\ddot{a}}{a} + \frac{\ddot{b}}{b} +
\frac{\dot{a}}{a} \frac{\dot{b}}{b} = -8 \pi G (p^I + p^A_{\|}), \\
\label{eq7}
&& 2 \, \frac{\ddot{a}}{a} + \left( \frac{\dot{a}}{a}
\right)^{\!2} = -8 \pi G (p^I + p^A_{\bot}).
\end{eqnarray}
In the following we shall restrict our analysis to the case of
matter-dominated universe ($p^I = 0$) filled with an anisotropic
component given by a uniform magnetic field (directed along the
$z$-axis), or a cosmic domain wall (whose plane of symmetry is the
$xy$-plane), or a cosmic string (directed along the $z$-axis).
\\
Magnetic fields have been observed on a wide range of scales. In
particular, they have been detected in galaxies, galaxy clusters,
and also in extra-galactic structures (for recent reviews on
cosmic magnetic fields, see Ref.~{\cite{CMF}}). It is reasonable
to assume that the actual observed magnetic fields have a
primordial origin. We assume that a (almost) uniform magnetic
field pervades our universe, though it is not excluded that such a
field may have a more complicated structure on small scales.
\\
Further examples of anisotropic components are given by cosmic
topological defects~\cite{Kibble}. As it is well known, phase
transitions in the early universe can generate domain walls
(cosmic strings), which are two-dimensional (one-dimensional)
defects that originate when a discrete (axial or cylindrical)
symmetry is broken.

For the three cases discussed above, the energy-momentum tensor
take on the form
\begin{equation}
\label{eq10}
(T_{B})^{\mu}_{\;\;\nu} = \rho_B \, \mbox{diag} (1,-1,-1,1),
\end{equation}
\begin{equation}
\label{eq18}
(T_{w})^{\mu}_{\;\;\nu} = \rho_w \, \mbox{diag} (1,1,1,0),
\end{equation}
\begin{equation}
\label{eq28}
(T_{s})^{\mu}_{\;\;\nu} = \rho_s \, \mbox{diag} (1,0,0,1),
\end{equation}
where $\rho_B$, $\rho_w$, and $\rho_s$, are the magnetic, wall,
and string energy density, respectively. Moreover, we assume that
the interaction of anisotropic components with matter is
negligible (in the case of magnetic fields, this corresponds to
taking into account that the conductivity of the primordial plasma
is very high~\cite{CMF}).
In this case, the anisotropic component of the energy-momentum
tensor is conserved, ${(T_A)}^{\mu}_{\;\; \nu;\mu} = 0$, so that
we have
\begin{equation}
\label{cons_anisotr}
{\dot{\rho}}^A + 2 \, \frac{\dot{a}}{a} (\rho^A + p^A_{\|}) +
\frac{\dot b}{b} \, ( \rho^A + p^A_{\bot}) = 0.
\end{equation}
Let us introduce the eccentricity
\begin{equation}
\label{eccentr}
e = \sqrt{1 - \left( \frac{b}{a} \right)^{\!\!2}} \, , \;\;\;\;
\mbox{or} \;\;\;\; e = \sqrt{1 - \left( \frac{a}{b}
\right)^{\!\!2}} \, ,
\end{equation}
and normalize the scale factors such that $a(t_0) = b(t_0) = 1$ at
the present time $t_0$. The first definition of eccentricity
applies to the cases of uniform magnetic field or string, while
the second one to the case of domain wall.
\\
In this paper, we restrict our analysis to the case of small
eccentricities (that is we consider the metric anisotropies as
perturbations over the isotropic Friedmann-Robertson-Walker
background).
In this limit, from Eqs.~(\ref{eq5})-(\ref{eq7}), we get the
following evolution equation for the eccentricity:
\begin{equation}
\label{evolution}
\frac{d (e \dot{e})}{dt} + 3 H (e \dot{e}) = \pm 8 \pi G(p^A_{\|}
- p^A_{\bot}),
\end{equation}
where the plus sign refers to the cases of magnetic field and
string, while the minus sign to the case of domain wall. Here $H =
\dot{a}/a$ is the usual Hubble expansion parameter for the
isotropic universe. In the matter-dominated era, it results $a(t)
\propto t^{2/3}$, so that $H = 2/(3t)$.
\\
Moreover, to the zero-order in the eccentricity, from
Eq.~(\ref{cons_anisotr}) it follows that the energy densities (and
pressure) scale in time as $\rho_B \propto a^{-4}$, $\rho_w
\propto a^{-1}$, and $\rho_s \propto a^{-2}$, for the three cases,
respectively.

The solution of Eq.~(\ref{evolution}) for the magnetic case is
\begin{equation}
\label{eq16}
e^2 = 8 \Omega_B^{(0)} (1 - 3 a^{-1} + 2 a^{-3/2}),
\end{equation}
where $\Omega_B^{(0)} = \rho_B(t_0)/\rho_{\rm cr}^{(0)}$, and
$\rho_{\rm cr}^{(0)} = 3 H_0^2/8 \pi G$ is the actual critical
energy density. At the decoupling, $t=t_{\rm dec}$, we have
$e_{\rm dec}^2 \simeq 16 \, \Omega_B^{(0)} z_{\rm dec}^{3/2}$,
where $e_{\rm dec} = e(t_{\rm dec})$ and $z_{\rm dec} \simeq 1088$
is the red-shift at decoupling~\cite{1yearWMAP}. Accordingly, we
get:
\begin{equation}
e_{\rm dec} \simeq 10^{-2} \left( \frac{\Omega_B^{(0)}}{10^{-7}}
\right)^{\!\!1/2} \! ,
\end{equation}
or
\begin{equation}
\label{eccentricity2}
e_{\rm dec} \simeq 10^{-2} h^{-1} \frac{B_0}{10^{-8} \G} \: ,
\end{equation}
where $B_0 = B(t_0)$ and $h \simeq 0.72$~\cite{1yearWMAP} is the
little-$h$ constant.
\\
If, for instance, we assume for the present cosmological magnetic
field strength the estimate $B_0 \simeq 5 \times 10^{-9} \G$,
which is compatible with the constraints analyzed in
Ref.~\cite{Ferreira}, we get an eccentricity at decoupling of
order $e_{\rm dec} \sim 10^{-2}$.

In the cases of domain wall and cosmic string, integrating
Eq.~(\ref{evolution}), we find
\begin{equation}
\label{eq26}
e^2 = \frac{2}{7} \, \Omega^{(0)}_w  (3a^2 + 4 a^{-3/2} - 7)
\end{equation}
and
\begin{equation}
\label{eq36}
e^2 = \frac{4}{5} \, \Omega_s^{(0)} (3a + 2a^{-3/2} -5),
\end{equation}
respectively, where $\Omega_w^{(0)}$ and $\Omega_s^{(0)}$ are the
actual energy densities, in units of $\rho_{\rm cr}^{(0)}$,
associated to domain wall and cosmic string.
\\
From the above equations, we can estimate the eccentricity at
decoupling in terms of wall and string energy densities at present
time:
\begin{equation}
e_{\rm dec} \simeq 10^{-2} \left( \frac{\Omega_w^{(0)}}{5 \times
10^{-7}} \right)^{\!\!1/2} \! ,
\end{equation}
and
\begin{equation}
e_{\rm dec} \simeq 10^{-2} \left( \frac{\Omega_s^{(0)}}{4 \times
10^{-7}} \right)^{\!\!1/2} \! .
\end{equation}
Vice versa, since the analysis of the CMB radiation constrains the
value of the eccentricity at decoupling to be less than $10^{-2}$
(see Ref.~\cite{prl} and next Section), one gets an upper limit on
the values of the energy density of cosmic defects stretching our
universe of order $\Omega_{w,s}^{(0)} \lesssim 10^{-7}$.

\section{\normalsize{CMB Quadrupole Anisotropy}}
\label{anisotropies}

\noindent Let us begin by briefly discussing the standard analysis
of the CMB temperature anisotropies~\cite{Dodelson}. First, the
temperature anisotropy is expanded in terms of spherical
harmonics:
\begin{equation}
\label{DeltaT}
\frac{\Delta T(\theta,\phi)}{\langle T \rangle} =
\sum_{l=1}^{\infty} \sum_{m=-l}^{l} a_{lm} Y_{lm}(\theta,\phi).
\end{equation}
After that, one introduces the power spectrum:
\begin{equation}
\label{spectrum}
\frac{\Delta T_l}{\langle T \rangle} = \sqrt{ \frac{1}{2 \pi} \,
\frac{l(l+1)}{2l+1} \sum_m |a_{lm}|^2},
\end{equation}
that fully characterizes the properties of the CMB anisotropy. In
particular, the quadrupole anisotropy refers to the multipole
$\ell=2$:
\begin{equation}
\label{quadrupole-T}
\mathcal{Q} \, \equiv \, \frac{\Delta T_2}{\langle T \rangle} \, ,
\end{equation}
where $\langle T \rangle \simeq 2.73$K is the actual (average)
temperature of the CMB radiation. The quadrupole problem resides
in the fact that the observed quadrupole anisotropy is in the
range (see Table 1):
\begin{equation}
\label{quad-obs}
\left(\Delta T_2 \right)_{\rm obs}^{2} \simeq (210 \div 276) \,
\mu \mbox{K}^2,
\end{equation}
while the expected quadrupole anisotropy according the
$\Lambda$CDM standard model is:
\begin{equation}
\label{quad-infl}
\left(\Delta T_2 \right)_{\rm I}^2 \simeq 1252 \, \mu \mbox{K}^2.
\end{equation}
If we admit that the large scale spatial geometry of our universe
is plane-symmetric with a small eccentricity, then we have that
the observed CMB anisotropy map is a linear superposition of two
contributions~\cite{prl,Bunn}:
\begin{equation}
\label{quad-sum}
\Delta T \; = \; \Delta T_{\rm A} + \, \Delta T_{\rm I} \; ,
\end{equation}
where $\Delta T_{\rm A}$ represents the temperature fluctuations
due to the anisotropic space-time background, while $\Delta T_{\rm
I}$ is the standard isotropic fluctuation caused by the
inflation-produced gravitational potential at the last scattering
surface. As a consequence, we may write:
\begin{equation}
\label{alm}
a_{lm} = a_{lm}^{\rm A} + \, a^{\rm I}_{lm}.
\end{equation}
We are interested in the distortion of the CMB radiation in a
universe with planar symmetry described by the metric (\ref{eq1}).
As before, we will work in the small eccentricity approximation.
From the null geodesic equation, we get that a photon emitted at
the last scattering surface having energy $E_{\rm dec}$ reaches
the observer with an energy equal to
$E_0(\widehat{n}) = \langle E_0 \rangle (1 - e_{\rm dec}^2
n_3^2/2)$,
where $\langle E_0 \rangle \equiv E_{\rm dec}/(1+z_{\rm dec})$,
and $\widehat{n} = (n_1,n_2,n_3)$ are the direction cosines of the
null geodesic in the symmetric (Robertson-Walker) metric.
\\
It is worth mentioning that the above result applies to the case
of the axis of symmetry  directed along the $z$-axis. We may, however,
easily generalize this result to the case where the symmetry axis
is directed along an arbitrary direction in a coordinate system
$(x_{\rm g},y_{\rm g},z_{\rm g})$ in which the $x_{\rm g} y_{\rm
g}$-plane is, indeed, the galactic plane. To this end, we perform
a rotation $\mathcal{R} = \mathcal{R}_{x}(\vartheta) \,
\mathcal{R}_{z}(\varphi + \pi/2)$ of the coordinate system
$(x,y,z)$, where $\mathcal{R}_{z}(\varphi + \pi/2)$ and
$\mathcal{R}_{x}(\vartheta)$ are rotations of angles $\varphi +
\pi/2$ and $\vartheta$ about the $z$- and $x$-axis, respectively.
In the new coordinate system the magnetic field is directed along
the direction defined by the polar angles $(\vartheta, \varphi)$.
Therefore, the temperature anisotropy in this new reference system
is:
\begin{equation}
\label{DeltaTA}
\frac{\Delta T_{\rm A}}{\langle T \rangle} \equiv \frac{E_0(n_{\rm
A}) -\langle E_0 \rangle}{\langle E_0 \rangle} = -\frac{1}{2} \,
e_{\rm dec}^2 n_{\rm A}^2 \; ,
\end{equation}
where $n_{\rm A} \equiv (\mathcal{R} \, \widehat{n})_3$ is equal
to
\begin{equation}
\label{nA}
n_{\rm A}(\theta,\phi) = \cos \theta \cos \vartheta - \sin \theta
\sin \vartheta \cos(\phi - \varphi) \; .
\end{equation}
Alternatively, when the eccentricity is small, Eq.~(\ref{eq1}) may
be written in a more standard form:
\begin{equation}
\label{metric2}
ds^2 = dt^2 - a^2(t) (\delta_{ij} + h_{ij}) \, dx^i dx^j,
\end{equation}
where $h_{ij}$ is the metric perturbation which takes on the form:
\begin{equation}
\label{metric-pert}
h_{ij} = -e^2 \delta_{i3} \delta_{j3}.
\end{equation}
The null geodesic equation in a perturbed
Friedmann-Robertson-Walker metric gives the temperature anisotropy
(Sachs-Wolfe effect):
\begin{equation}
\label{anis2}
\frac{\Delta T}{\langle T \rangle}  =  -\frac{1}{2} \int_{t_{\rm
dec}}^{t_0} \! dt \; \frac{\partial h_{ij}}{\partial t} \; n^i
n^j,
\end{equation}
where $n^i$ are the directional cosines. Using $e(t_0)=0$, from
Eqs.~(\ref{metric-pert}) and (\ref{anis2}) one gets:
\begin{equation}
\label{anis3}
\frac{\Delta T}{\langle T \rangle}  =  - \frac{1}{2} \,  e^2_{\rm
dec}n_3^2,
\end{equation}
which indeed agrees with our previous result. \\
%
%
%
%
%
%
It is easy to see from Eq.~(\ref{DeltaTA}) that only the
quadrupole terms ($\ell = 2$) are different from zero:
\begin{eqnarray}
\label{almA}
&& a_{20}^{\rm A} = -\frac{\sqrt{\pi}}{6\sqrt{5}} \,
                  [1 + 3\cos(2 \vartheta) ] \, e_{\rm dec}^2 \, , \nonumber \\
&& a_{21}^{\rm A} = -(a_{2,-1}^{\rm A})^{*} =
                  \sqrt{\frac{\pi}{30}} \;
                  e^{-i \varphi}  \sin(2\vartheta) \, e_{\rm dec}^2 \, , \nonumber \\
&& a_{22}^{\rm A} = (a_{2,-2}^{\rm A})^{*} = -
                  \sqrt{\frac{\pi}{30}}
                  \; e^{-2 i \varphi} \sin^2\!\vartheta \,
                  e_{\rm dec}^2 \, .
\end{eqnarray}
Consequently, the quadrupole anisotropy is:
\begin{equation}
\mathcal{Q}_{\rm A} = \frac{2}{5 \sqrt{3}} \; e_{\text dec}^2 \, .
\end{equation}
Since the temperature anisotropy is a real function, we have
$a_{l,-m} = (-1)^m (a_{l,m})^*$. Observing that $a_{l,-m}^{\rm A}
= (-1)^m (a_{l,m}^{\rm A})^*$ [see Eq.~(\ref{almA})], we get
$a^{\rm I}_{l,-m} = (-1)^m (a^{\rm I}_{l,m})^*$.
Moreover, because the standard inflation-produced temperature
fluctuations are statistically isotropic, we will make the
reasonable assumption that the $a^{\rm I}_{2m}$ coefficients are
equals up to a phase factor. Therefore, we can write:
\begin{eqnarray}
\label{almI}
&& a^{\rm I}_{20} = \sqrt{\frac{\pi}{3}} \; e^{i \phi_1} \mathcal{Q}_{\rm I}, \nonumber \\
&& a^{\rm I}_{21} =  - (a^{\rm I}_{2,-1})^{*} = \sqrt{\frac{\pi}{3}} \; e^{i \phi_2} \mathcal{Q}_{\rm I} \; , \\
&& a^{\rm I}_{22} =  (a^{\rm I}_{2,-2})^{*} = \sqrt{\frac{\pi}{3}}
\; e^{i \phi_3} \mathcal{Q}_{\rm I} \; , \nonumber
\end{eqnarray}
where $0 \leq \phi_i \leq 2 \pi$ are unknown phases, and
\begin{equation}
\label{QInflation}
\mathcal{Q}_{\rm I} \simeq 13 \times 10^{-6}.
\end{equation}
Taking into account Eqs.~(\ref{almA}) and (\ref{almI}), and
Eqs.~(\ref{DeltaT}), ~(\ref{spectrum}) and~(\ref{alm}), we get for
the total quadrupole:
\begin{equation}
\label{quadrupole}
\mathcal{Q}^2 = \mathcal{Q}_{\rm A}^2 + \mathcal{Q}_{\rm I}^2 - 2f
\mathcal{Q}_{\rm A} \mathcal{Q}_{\rm I},
\end{equation}
where
\begin{eqnarray}
\label{f}
f(\vartheta, \varphi \, ; \phi_1, \phi_2, \phi_3) & = &
             \frac{1}{4\sqrt{5}} \,
             \{ 2 \sqrt{6} \, \left[\sin\vartheta \cos(2\varphi + \phi_3) -
             2 \cos\vartheta \cos(\varphi + \phi_2) \right] \sin\vartheta \nonumber \\
             & + & [1 + 3 \cos(2 \vartheta) ] \cos\phi_1 \}.
\end{eqnarray}
Looking at Eq.~(\ref{quadrupole}) we see that, if the space-time
background is not isotropic, the quadrupole anisotropy can become
smaller than the one expected in the standard picture of the
$\Lambda$CDM (isotropic-) cosmological model of temperature
fluctuations. We may fix the direction of the magnetic field and
the eccentricity by solving Eq.~(\ref{alm}), which is (for $\ell=2$)
a system of 5 equations containing 5 unknown parameters: $e_{\text
dec}$, $\vartheta$, $\varphi$, $\phi_2$, and $\phi_3$. Note that it is
always possible to choose $a_{20}$ real, and then $\phi_1 = 0$. \\
\indent
To solve Eq.~(\ref{alm}) for $\ell=2$, we need the observed values
of the $a^{\rm}_{2m}$'s. We use the recent cleaned CMB temperature
fluctuation map of the WMAP3 data obtained by using an improved
internal linear combination method as Galactic foreground
subtraction technique. In particular, we adopt the three maps
SILC400~\cite{SILC400}, WILC3YR~\cite{WILC3YR}, and
TCM3YR~\cite{TCM3YR}. For completeness, we report in Table 1 the
values of $a^{\rm}_{2m}$ corresponding to these maps. We also
indicate the values of $\phi_2$ and $\phi_3$ [found by solving
Eq.~(\ref{almI})] in the case of isotropic universe ($a_{2m}^{\rm
A}=0$).
Moreover, for sake of definiteness, we assume that the planar
symmetry is induced by a cosmological magnetic field with strength
$B_0$ at the present time.


\begin{table}
\begin{center}
\caption{The cleaned maps SILC400, WILC3YR, and TCM3YR. Note that
the values of $a_{2m}$ in this table correspond to the values of
$a_{2m}$ given in Refs.~\cite{SILC400,WILC3YR,TCM3YR} divided by
$\langle T \rangle \simeq 2.73 K$. The values of the angles
$\phi_2$, and $\phi_3$ are in degrees.} \vspace{0.5cm}
\begin{tabular}{llllllll}

\hline \hline

&MAP     &$m$ &$\mbox{Re}[a_{2m}]/10^{-6}$ &$\mbox{Im}[a_{2m}]/10^{-6}$ &$~~\phi_2$ &$~\,\phi_3$ &$\left(\Delta T_2\right)^2/\mu\mbox{K}^2$ \\
\hline
&        &$0$ &~~~~\:$2.75$                &~~~~\:$0.00$                &           &            & \\
&SILC400 &$1$ &~~~\!$-0.56$                &~~~~\:$1.77$                &$107.6$    &$44.2$      &~~~~\:$275.8$ \\
&        &$2$ &~~~\!$-6.79$                &~~~\!$-6.60$                &           &            & \\
\hline
&        &$0$ &~~~~\:$4.21$                &~~~~\:$0.00$                &           &            & \\
&WILC3YR &$1$ &~~~\!$-0.02$                &~~~~\:$1.78$                &$90.6$     &$52.5$      &~~~~\:$248.8$ \\
&        &$2$ &~~~\!$-5.28$                &~~~\!$-6.89$                &           &            & \\
\hline
&        &$0$ &~~~~\:$1.22$                &~~~~\:$0.00$                &           &            & \\
&TCM3YR  &$1$ &~~~~\:$0.10$                &~~~~\:$1.79$                &$86.8$     &$49.2$      &~~~~\:$209.5$ \\
&        &$2$ &~~~\!$-5.45$                &~~~\!$-6.32$                &           &            & \\

\hline \hline

\end{tabular}
\end{center}
\end{table}



\begin{table}
\begin{center}
\caption{Numerical solutions of Eq.~(\ref{alm}) obtained by using
the map SILC400; the values of the angles $\vartheta$, $\varphi$,
$\phi_2$, and $\phi_3$ are in degrees.} \vspace{0.5cm}
\begin{tabular}{lllllll}

\hline \hline

&$e_{\dec}/10^{-2}$ &$~~~\vartheta$ &$~~~~~\varphi$ &$~~~\phi_2$ &$~~\,\phi_3$ &$~~B_0/10^{-9}\G$ \\

\hline

&$~~0.69$           &$~38.6$        &$~~~80.1$      &$~100.8$    &$~34.8$      &$~~~~~~~5.0$  \\
&$~~0.69$           &$~38.4$        &$~~~99.2$      &$~83.7$     &$~20.9$      &$~~~~~~~4.9$  \\
&$~~0.68$           &$~38.3$        &$~~~37.2$      &$~139.1$    &$~66.8$      &$~~~~~~~4.9$  \\
&$~~0.67$           &$~37.4$        &$~~~139.3$     &$~47.3$     &$~12.8$      &$~~~~~~~4.8$  \\
&$~~0.63$           &$~35.3$        &$~~~308.9$     &$~43.2$     &$~17.3$      &$~~~~~~~4.5$  \\
&$~~0.62$           &$~34.5$        &$~~~281.2$     &$~73.6$     &$~25.8$      &$~~~~~~~4.5$  \\
&$~~0.62$           &$~34.3$        &$~~~239.8$     &$~122.5$    &$~48.9$      &$~~~~~~~4.4$  \\
&$~~0.62$           &$~34.3$        &$~~~255.2$     &$~104.2$    &$~39.9$      &$~~~~~~~4.4$  \\

\hline \hline

\end{tabular}
\end{center}
\end{table}



\begin{table}
\begin{center}
\caption{Numerical solutions of Eq.~(\ref{alm}) obtained by using
the map WILC3YR; the values of the angles $\vartheta$, $\varphi$,
$\phi_2$, and $\phi_3$ are in degrees.} \vspace{0.5cm}
\begin{tabular}{lllllll}

\hline \hline

&$e_{\dec}/10^{-2}$ &$~~~\vartheta$ &$~~~~~\varphi$ &$~~~\phi_2$ &$~~\,\phi_3$ &$~~B_0/10^{-9}\G$ \\

\hline

&$~~0.69$           &$~40.6$        &$~~~91.7$      &$~88.6$     &$~28.0$      &$~~~~~~~4.9$ \\
&$~~0.69$           &$~40.6$        &$~~~88.3$      &$~91.6$     &$~30.8$      &$~~~~~~~4.9$ \\
&$~~0.67$           &$~40.0$        &$~~~139.4$     &$~45.6$     &$~8.6$       &$~~~~~~~4.8$ \\
&$~~0.67$           &$~40.0$        &$~~~40.4$      &$~134.6$    &$~71.0$      &$~~~~~~~4.8$ \\

\hline \hline

\end{tabular}
\end{center}
\end{table}



\begin{table}
\begin{center}
\caption{Numerical solutions of Eq.~(\ref{alm}) obtained by using
the map TCM3YR; the values of the angles $\vartheta$, $\varphi$,
$\phi_2$, and $\phi_3$ are in degrees.} \vspace{0.5cm}
\begin{tabular}{lllllll}

\hline \hline

&$e_{\dec}/10^{-2}$ &$~~~\vartheta$ &$~~~~~\varphi$ &$~~~\phi_2$ &$~~\,\phi_3$ &$~~B_0/10^{-9}\G$ \\

\hline

&$~~0.70$           &$~36.5$        &$~~~96.6$      &$~83.8$      &$~24.9$     &$~~~~~~~5.0$ \\
&$~~0.70$           &$~36.5$        &$~~~83.2$      &$~95.7$      &$~35.1$     &$~~~~~~~5.0$ \\
&$~~0.69$           &$~36.1$        &$~~~129.4$     &$~54.3$      &$~9.4$      &$~~~~~~~5.0$ \\
&$~~0.69$           &$~36.0$        &$~~~51.6$      &$~124.1$     &$~59.8$     &$~~~~~~~5.0$ \\

\hline \hline

\end{tabular}
\end{center}
\end{table}


Numerical solutions of Eq.~(\ref{alm}), referring to the three
maps, are given in Tables 2, 3, and 4, respectively.
In the Appendix, we will show that the system (\ref{alm}) admits
at most 8 independent solutions. However, due to the particular
values taken by the $a_{2m}$ for the three different maps, we find
that to the map SILC400 it corresponds 8 independent solutions,
while to the maps WILC3YR and TCM3YR there correspond only 4
independent solutions.
\\
Moreover we observe that, for each independent solution $(e_{\rm
dec},\vartheta,\varphi,\phi_2,\phi_3)$ shown in the tables, there
exists another one given by $(e_{\rm dec},\pi-\vartheta,\varphi
\pm \pi,\phi_2,\phi_3)$, where we take the plus sign if $\varphi <
\pi$ and the minus sign if $\varphi > \pi$.

Looking at Table 1 and Eq.~(\ref{QInflation}), we see that the
values of coefficients $|a_{21}|$ are about one order of magnitude
smaller then $\mathcal{Q}_{\rm I}$. Assuming $\mathcal{Q}_{\rm I}
\gg |a_{21}|$, we will show in the Appendix that an approximate
solution for $e_{\rm dec}$ and $\vartheta$ is:
\begin{equation}
\label{eapprox}
e_{\rm dec}^2 \simeq \sqrt{10} \, \mathcal{Q}_{\rm I},
\end{equation}
and
\begin{equation}
\label{tetaapprox}
\vartheta \simeq \frac{1}{2} \arccos \! \left[
\frac{\left(5\sqrt{5} - \sqrt{6\pi}\,\right) \!\! \mathcal{Q}_{\rm
I} - 5\sqrt{3} \, a_{20}}{3\sqrt{6\pi} \mathcal{Q}_{\rm I}}
\right] \!.
\end{equation}
From Eq.~(\ref{eapprox}) we also have that $e_{\rm dec} \simeq
0.64 \times 10^{-2}$, $B_0 \simeq 4.6 \times 10^{-9} \G$, and
$\mathcal{Q}_{\rm A} \simeq (4/\sqrt{30}) \mathcal{Q}_{\rm I}
\simeq 0.7 \mathcal{Q}_{\rm I}$. From Eq.~(\ref{tetaapprox}) we
get, for the three different maps, $\vartheta \simeq
34^{\circ},36^{\circ},31^{\circ}$, respectively. As one can easily
check, the approximate solutions Eqs.~(\ref{eapprox}) and
(\ref{tetaapprox}) are quite close to the numerical values.


\begin{figure}[h]
\begin{center}
\includegraphics[clip,width=0.7\textwidth]{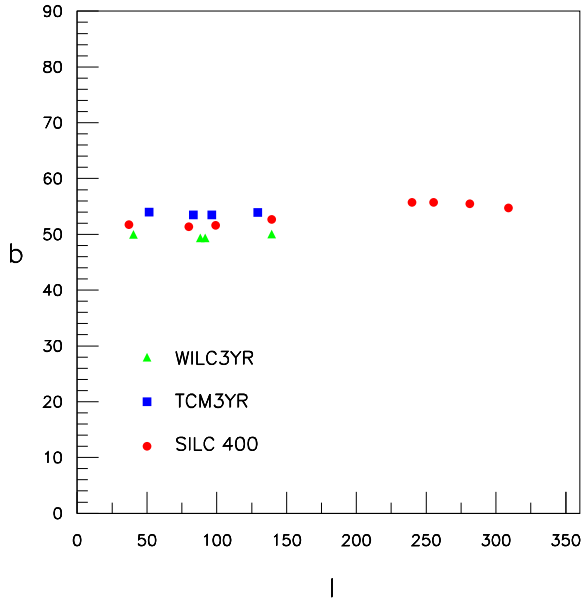}
\caption{Numerical solutions of Eq.~(\ref{alm}) obtained by using
the three maps. Note that $b=90^{\circ}-\vartheta$, and
$l=\varphi$, where $(b,l)$ are the galactic coordinates.}
\end{center}
\end{figure}


In Fig.~1, we plot the numerical solutions $\vartheta$ versus
$\varphi$ (which define the direction of the symmetry axis), using
the so-called galactic coordinates system characterized by the
galactic latitude $b$, and galactic longitude $l$. In our
notation, the angle $b$ corresponds to $b = 90^{\circ} -
\vartheta$ while $l = \varphi$. From Fig.~1 we see that the
galactic latitude of the symmetry axis is remarkably independent
on the adopted CMB temperature fluctuation map. On the contrary,
we find that the galactic longitude is poorly constrained, for it
may vary in a large interval. Indeed, we have $b \simeq 50^{\circ}
- 54^{\circ}$, while $40^{\circ} \lesssim l \lesssim 140^{\circ}$,
$240^{\circ} \lesssim l \lesssim 310^{\circ}$.
One could think of choosing the value of $\varphi$ (for each map)
by imposing that the corresponding solution $(e_{\rm
dec},\vartheta,\varphi,\phi_2,\phi_3)$ has the smallest difference
between $(\phi_2,\phi_3)$ and those in absence of anisotropic
component listed in Table 1. Since such a difference indicates how
much the intrinsic quadrupole needs to be rotated to accommodate
the anisotropic component, the above choice would represent the
``least radical modification'' to the standard (isotropic)
cosmological model. For the three maps SILC400, WILC3YR, TCM3YR,
we could then pick the solutions corresponding to $l = \varphi
\simeq 255^{\circ}, 88^{\circ}, 83^{\circ}$, respectively. Also in
this case, however, we cannot univocally fix the galactic
longitude of the magnetic field.

It is interesting to stress that our determination of the
direction of the symmetry axis is in fair agreement with the
statistical analysis of Ref.~\cite{SILC400} which confirmed the
strong alignment and planarity of the quadrupole and the octupole.
Thus, we see that our proposal for an ellipsoidal universe could
provide a natural solution to the low-$\ell$ CMB puzzles.

\section{\normalsize{Limits on Cosmic Magnetic Fields from Big Bang
Nucleosynthesis and Large Scale Structure formation.}}

\noindent In the previous Section, we have seen that ellipsoidal
universe with an eccentricity at decoupling $e_{\rm dec} \simeq
0.64 \times 10^{-2}$ could provide a natural solution to the low
quadrupole CMB puzzle. We believe that the most interesting and
intriguing possibility is plane-symmetric geometry induced by
cosmological magnetic fields. In fact, magnetic fields are
observed in the universe up to cosmological scales. So that, we
cannot exclude that the origin of the presently-observed cosmic
magnetic fields is primordial as long the predictions of the
standard cosmological model are not invalidate. Indeed, our
estimate for the present-time magnetic field strength,
\begin{equation}
\label{eq5.1}
B_0 \; \simeq \; 4.6 \times 10^{-9}\;  \G,
\end{equation}
is in agreement with the observed magnetic fields within galaxies
and clusters of galaxy~\cite{CMF}. However, the presence of a
cosmic primordial magnetic field must also fulfil the constraints
coming from Big Bang Nucleosynthesis (BBN) and Large Scale
Structure (LSS) formation.
\\
\\
{\it Limits from BBN.} Since in the early universe the
conductivity of the primordial plasma is very high, magnetic
fields are frozen into the plasma and evolve adiabatically, $B
\propto a^{-2}$, where $a \propto g_{*S}^{-1/3} T^{-1}$, $T$ being
the temperature, and $g_{*S}$ counts the total number of
effectively massless degrees of freedom referring
to the entropy density of the universe~\cite{Kolb}. \\
\indent
The limit coming from BBN refers to uniform magnetic fields at
that time. The upper bound is given in Ref.~\cite{Grasso}:
\begin{equation}
\label{limitBBN1}
B(T_{\BBN}) \lesssim 1 \times 10^{11} \G,
\end{equation}
where $T_{\BBN} = 10^9 \mbox{K} \simeq 0.1 \MeV$. This limit
translates into:
\begin{equation}
\label{limitBBN2}
B_0 =
\left(\frac{g_{*S}({T_0})}{g_{*S}({T_{\BBN}})}\right)^{\!2/3}
\left(\frac{T_0}{T_{\BBN}}\right)^{\!2} \, B(T_{\BBN}) \lesssim 6
\times 10^{-7} \G,
\end{equation}
where $B_0 = B(T_0)$, and we used $g_{*S}({T_{\BBN}}) \simeq
g_{*S}({T_0}) \simeq 3.91$, and $T_0 \simeq 2.35 \times 10^{-4}
\eV$ \cite{Kolb}. We see, indeed, that our estimate~(\ref{eq5.1})
does not violate the upper bound~(\ref{limitBBN2}).
\\
\\
{\it Limits from LSS.} Anisotropic cosmological models have been
extensively studied since long time (see, for instance,
Ref.~\cite{Zeldovich} and references therein). It is known that
strong anisotropic expansion could influence the formation of
large-scale structures. In particular, in case of strong
anisotropies in the expansion of the universe the time of growth
of structures could increases by a factor $3 - 5$~\cite{Zeldovich}. \\
In matter-dominated universe, we see from Eq.~(\ref{eq16}) that
the eccentricity evolves as $e^2(z) \propto z^{3/2}$. So that, at
the epoch of matter-radiation equality we get:
\begin{equation}
e_{\eq} \simeq \left( \frac{z_{\eq}}{z_{\dec}} \right)^{\!3/2}
e_{\dec} \simeq 5.5 \, e_{\dec} \simeq 3.5 \times 10^{-2},
\end{equation}
where $e_{\eq} = e(z_{\eq})$ and we used $z_{\dec} \simeq 1088$,
$z_{\eq} \simeq 3400$ \cite{Kolb}, and $e_{\dec} \simeq  0.64
\times 10^{-2}$. Thus, we see that at the epoch of
matter-radiation equality the anisotropy in the cosmological
expansion is small. So that, we do not expect dramatic effects on
the processes of growth of structures. Nevertheless, it could well
be that small anisotropies in the cosmic expansion could leave
signatures on large scales which should be contrasted with
observations.

\section{\normalsize{Conclusions}}

The recent measurements of the cosmic microwave background angular
spectrum have greatly improved our understanding of the universe,
in particular we have information on the properties and origin of
the density fluctuation of the cosmic plasma. In fact, the
observed CMB fluctuations are in remarkable agreement with the
prediction of the $\Lambda$CDM standard model with scale-invariant
adiabatic fluctuations generated during the inflationary epoch.
However, some anomalies have been discovered on large angular
scales including, in particular, the low CMB quadrupole power and
the alignment and planarity of the quadrupole and octupole modes.
It should be stressed, however, that the large-angle anomalies in
the CMB anisotropy are still subject to an intense
debate~\cite{AnomalyDebate}. For instance, it has been suggested
that the CMB anomalies could be explained by the residual Galactic
foreground emission~\cite{Rakic}. Therefore, it is very important
to improve and develop techniques which allow a
better removal of residual foreground contamination~\cite{Foreground}. \\
\indent
In this paper, we have further elaborated our previous
proposal~\cite{prl} that an ``ellipsoidal expansion'' of the
universe could resolve the CMB quadrupole anomaly. We have shown
that such anisotropic expansion (described by a plane-symmetric
metric) can be generated by cosmological magnetic fields or
topological defects, such as cosmic domain walls or cosmic
strings. Indeed, topological cosmic defects are relic structures
that are predicted to be produced in the course of symmetry
breaking in the hot, early universe. Nevertheless, we believe that
the most interesting and intriguing possibility is plane-symmetric
geometry induced by cosmological magnetic fields. In fact,
magnetic fields have been already observed in the
universe up to cosmological scales. \\
\indent
We have shown that the quadrupole anomaly can be resolved if the
last scattering surface of CMB is an ellipsoid. Indeed, we found
that, if the eccentricity at decoupling is:
\begin{equation}
\label{eqc1} e_{\rm dec} \simeq  0.64 \times 10^{-2},
\end{equation}
then the quadrupole amplitude can be drastically reduced without
affecting higher multipoles of the angular power spectrum of the
temperature anisotropy. Remarkably, our estimate of $e_{\rm dec}$
gives for the strength of the cosmic magnetic field:
\begin{equation}
\label{eqc2}
B_0 \; \simeq \; 4.6 \times 10^{-9} \;  \G,
\end{equation}
which agrees with the limits arising from primordial
nucleosynthesis and large scale structure formation. Moreover, we
have obtained tight constraints on the direction $(b,l)$ of the
axis of symmetry:
\begin{equation}
\label{eqc3}
b \; \simeq \; 50^{\circ}\;  - \;  54^{\circ}, \; \;\; 40^{\circ}
\; \lesssim \;  l  \; \lesssim \; 140^{\circ}, \;\;\; 240^{\circ}
\; \lesssim \; l \; \lesssim \; 310^{\circ},
\end{equation}
where $b$ and $l$ are the galactic latitude and the galactic
longitude, respectively.
This constraints are in fair agreement with recent statistical
analyses of the cleaned CMB temperature fluctuation maps of the
WMAP3 data obtained by using an improved internal linear
combination method as Galactic foreground subtraction technique. \\
\indent
In conclusion, our proposal for the ellipsoidal universe offers a
natural solution to the CMB quadrupole anomaly which future
pattern searches with more refined data, such as further WMAP data
releases or PLANK data, will be able to confirm or reject. In
addition, recently it has been shown~\cite{Cea} that the large
scale polarization of the cosmic microwave background induced by
an ellipsoidal universe compares quite well to the average level
of polarization detected by the Wilkinson Microwave Anisotropy
Probe. Still, it is necessary to better understand the foreground
contamination of the polarization to reach a firm conclusion.
\\
Finally, we find amusing that there are already independent
indications of a symmetry axis in the large-scale geometry of the
universe, coming from the analysis of spiral galaxies in the Sloan
Digital Sky Survey~\cite{Longo} and the analysis of polarization
of electromagnetic radiation propagating over cosmological
distances~\cite{Nodland}.

\newpage

\section{\normalsize{Appendix}}

\noindent In this Appendix, we solve the system of equations
(\ref{alm}) for $\ell=2$. The numerical values of the $a_{2m}$'s
are listed in Table 1, while the $a_{2m}^{\rm A}$'s and $a^{\rm
I}_{2m}$'s are given by Eq.~(\ref{almA}) and Eq.~(\ref{almI}),
respectively. \\
\indent
Since it is always possible to choose $a_{20}$ real, we take
$\phi_1 = 0$. Moreover, the temperature anisotropy is real so that
we have $a_{2,-m} = (-1)^m (a_{2m})^*$, $a_{2,-m}^{\rm A} = (-1)^m
(a_{2m}^{\rm A})^*$, and $a^{\rm I}_{2,-m} = (-1)^m (a^{\rm
I}_{2m})^*$.
Therefore, the system of equations (\ref{alm}) reduces to:
\begin{eqnarray}
\label{system1}
&& a_{20} = -\frac{\sqrt{\pi}}{6\sqrt{5}} \, [1 +
            3\cos(2\vartheta) ] \,
            e_{\rm dec}^2 \, + \,
            \sqrt{\frac{\pi}{3}} \, \mathcal{Q}_{\rm I}, \\
\label{system2}
&& \mbox{Re}[a_{21}] = \sqrt{\frac{\pi}{30}} \,
                       \cos\!\varphi \, \sin(2\vartheta) \, e_{\rm dec}^2 \, + \,
                       \sqrt{\frac{\pi}{3}} \, \cos\!\phi_2 \mathcal{Q}_{\rm I}, \\
\label{system3}
&& \mbox{Im}[a_{21}] = -\sqrt{\frac{\pi}{30}} \,
                       \sin\!\varphi \, \sin(2\vartheta) \, e_{\rm dec}^2 \, + \,
                       \sqrt{\frac{\pi}{3}} \, \sin\!\phi_2 \mathcal{Q}_{\rm I}, \\
\label{system4}
&& \mbox{Re}[a_{22}] = -\sqrt{\frac{\pi}{30}} \,
                       \cos\!\varphi \, \sin^2\!\vartheta \, e_{\rm dec}^2 \, + \,
                       \sqrt{\frac{\pi}{3}} \, \cos\!\phi_3 \mathcal{Q}_{\rm I}, \\
\label{system5}
&& \mbox{Im}[a_{22}] = \sqrt{\frac{\pi}{30}} \,
                       \sin\!\varphi \, \sin^2\!\vartheta \, e_{\rm dec}^2 \, + \,
                       \sqrt{\frac{\pi}{3}} \, \sin\!\phi_3 \mathcal{Q}_{\rm I},
\end{eqnarray}
where $\mathcal{Q}_{\rm I}$ is given by Eq.~(\ref{QInflation}).
We see that these  equations form a system of 5 transcendental equations
containing 5 unknown parameters: $e_{\rm dec}$, $\vartheta$,
$\varphi$, $\phi_2$, and $\phi_3$. Solving Eq.~(\ref{system1})
with respect to $\vartheta$ we get two independent solutions:
\begin{equation}
\label{tetatilde0}
\vartheta = \{ \widetilde{\vartheta}, \pi -\widetilde{\vartheta} \},
\end{equation}
where
\begin{equation}
\label{tetatilde}
\widetilde{\vartheta} = \frac{1}{2} \arccos \! \left(
\frac{5\sqrt{5\pi}\mathcal{Q}_{\rm I} - 5\sqrt{15} \, a_{20} -
\sqrt{3\pi} \, e_{\rm dec}^2}{3\sqrt{3\pi} \, e_{\rm dec}^2}
\right) \! .
\end{equation}
Squaring Eqs.~(\ref{system4}) and (\ref{system5}), adding side by
side, and then solving with respect to $\varphi$, we obtain 8
independent solutions:
\begin{equation}
\label{phitilde0}
\varphi = \{ \widetilde{\varphi}_{\pm}, \pi
-\widetilde{\varphi}_{\pm}, 2\pi -\widetilde{\varphi}_{\pm}, \pi +
\widetilde{\varphi}_{\pm} \},
\end{equation}
where
\begin{equation}
\label{phitilde}
\widetilde{\varphi}_{\pm} = \frac{1}{2} \arccos \! \left(
\frac{\alpha \gamma \pm \beta \sqrt{\alpha^2 + \beta^2 -
\gamma^2}}{\alpha^2 + \beta^2} \right) \!,
\end{equation}
and
\begin{eqnarray}
\label{abc}
\alpha \!\!&=&\!\!
\sqrt{\frac{2\pi}{15}} \: \mbox{Re}[a_{22}] \sin^2\!\vartheta  \, e_{\rm dec}^2 \, , \\
\beta \!\!&=&\!\!
\sqrt{\frac{2\pi}{15}} \: \mbox{Im}[a_{22}] \sin^2\!\vartheta \, e_{\rm dec}^2 \, , \\
\gamma \!\!&=&\!\! \frac{\pi}{30} \, \sin^4 \!\vartheta \, e_{\rm
dec}^4 + |a_{22}|^2 - \frac{\pi}{3} \mathcal{Q}_{\rm I}^2.
\end{eqnarray}
By dividing side by side Eqs.~(\ref{system3}) and (\ref{system2}),
and solving with respect to $\phi_2$, we get:
\begin{equation}
\label{phi2}
\tan\!\phi_2 = \frac{\sqrt{30} \, \mbox{Im}[a_{21}] + \sqrt{\pi}
\sin\!\varphi \, \sin(2\vartheta) \, e_{\rm dec}^2}{\sqrt{30} \,
\mbox{Re}[a_{21}] - \sqrt{\pi} \sin\!\varphi \, \sin(2\vartheta)
\, e_{\rm dec}^2} \, .
\end{equation}
The same procedure applied to Eqs.~(\ref{system5}) and
(\ref{system3}) results in:
\begin{equation}
\label{phi3}
\tan\!\phi_3 = \frac{\sqrt{30} \, \mbox{Im}[a_{22}] - \sqrt{\pi}
\sin(2\varphi) \, \sin^2\!\vartheta \, e_{\rm dec}^2}{\sqrt{30} \,
\mbox{Re}[a_{22}] + \sqrt{\pi} \sin(2\varphi) \, \sin^2\!\vartheta
\, e_{\rm dec}^2} \, .
\end{equation}
Finally, by squaring Eqs.~(\ref{system2}) and (\ref{system3}), and adding side
by side, we get:
\begin{equation}
\label{eequation}
e_{\rm dec}^4 + 2c \, e_{\rm dec}^2 - d = 0 \; ,
\end{equation}
where we have defined
\begin{eqnarray}
\label{cd1}
c(\varphi,\vartheta) \!\!&=&\!\! \sqrt{\frac{30}{\pi}} \,
(\mbox{Re}[a_{21}] \cos\varphi - \mbox{Im}[a_{21}] \sin\varphi) \csc(2\vartheta), \\
\label{cd2}
d(\varphi,\vartheta) \!\!&=&\!\! 10 \! \left( \mathcal{Q}_{\rm
I}^2 - \frac{3}{\pi} \, |a_{21}|^2  \right) \!
\csc^2\!(2\vartheta) \; .
\end{eqnarray}
We observe that the couple $(\vartheta, \varphi)$ can assume 16
different values, according to Eqs.~(\ref{tetatilde0}) and
(\ref{phitilde0}). Inserting these values in
Eqs.~(\ref{eequation})-(\ref{cd2}) we arrive at 16 different
equations for $e_{\rm dec}$. It is straightforward to verify that
only 8 of these are ``independent'', in the sense that, given a
solution $(e_{\rm dec},\vartheta,\varphi)$ of one of the
``independent'' equations, then $(e_{\rm dec}, \pi-\vartheta,
\varphi \pm \pi)$  is a solutions of one of the ``dependent'' ones
(we must  take the plus sign if $\varphi < \pi$ and the minus sign
if $\varphi > \pi$). The 8 independent equations can be solved
numerically and their solutions are presented in Table 2, 3, and
4. Here, we just observe that, to the map SILC400 it corresponds 8
independent solutions, while to the maps WILC3YR and TCM3YR there
correspond only 4 independent solutions. \\
\indent
Finally, we derive the approximate solutions
(\ref{eapprox})-(\ref{tetaapprox}). To this end, we may formally solve
Eq.~(\ref{eequation}) to get $e_{\rm dec}^2 = c \pm \sqrt{c^2 +
d}$.
If $\mathcal{Q}_{\rm I} \gg |a_{21}|$ then $d \gg c^2$, and we
obtain $e_{\rm dec}^2 \simeq \sqrt{10} \, \mathcal{Q}_{\rm I}\,
|\csc(2\vartheta)|$.
Now, it is easy to check numerically that $|\csc(2\vartheta)|
\simeq 1$. Indeed, from Table 2, 3, 4, we get $1.01 \lesssim
|\csc(2\vartheta)| \lesssim 1.07$. As a consequence, we have
$e_{\rm dec}^2 \simeq \sqrt{10} \, \mathcal{Q}_{\rm I}$, which
indeed agrees with Eq.~(\ref{eapprox}).
Inserting Eq.~(\ref{eapprox}) into Eq.~(\ref{tetatilde}) we easily
recover Eq.~(\ref{tetaapprox}).

\end{document}